# Identifying Oriented Spin Space Groups and Related Physical Properties Using an Online Platform FINDSPINGROUP

Yutong Yu[1], Xiaobing Chen[2,1,†] Yanzhou Zhu[3], Yuhui Li[1], Renzheng Xiong[1], Jiayu Li[1,4], Yuntian Liu[1], and Qihang Liu[1,2,5,*]

[1]*State Key Laboratory of Quantum Functional Materials, Department of Physics, and Guangdong Basic Research Center of Excellence for Quantum Science, Southern University of Science and Technology (SUSTech), Shenzhen 518055, China*

[2]*Quantum Science Center of Guangdong–Hong Kong–Macao Greater Bay Area (Guangdong), Shenzhen 518045, China*

[3]*School of Mathematics Science, Xiamen University, Xiamen 361005, China*

[4]*New Cornerstone Science Lab, Department of Physics, The University of Hong Kong, Hong Kong, China*

[5]*Guangdong Provincial Key Laboratory for Computational Science and Material Design, Southern University of Science and Technology, Shenzhen 518055, China*

[†]Email: chenxb@sustech.edu.cn

[*]Email: liuqh@sustech.edu.cn

## Abstract

Unconventional magnets that combine antiferromagnetic structures with ferromagnetic-like responses are essential for the development of next-generation spintronics. Their emergent properties are fundamentally dictated by the interplay between exchange-driven magnetic geometry and spin-orbit coupling, which are described by spin space group (SSG) and magnetic space group (MSG) frameworks, respectively. However, the lack of direct correspondence between these frameworks, developed in different eras, hinders the systematic tracking of symmetry evolution of these intertwined physical contributions. In this work, we introduce FINDSPINGROUP, a computational architecture that implements the recently emerged, oriented spin space group framework to unify SSG and MSG descriptions. By automating the tracking of symmetry-breaking pathways from the non-relativistic to the relativistic limit, this online platform enables the classification of magnetic phases and the derivation of symmetry-constrained tensors for phenomena such as momentum-dependent spin splitting and the anomalous Hall effect. By standardizing data exchange through the spin crystallographic information file, this architecture establishes a computational infrastructure for the high-throughput discovery and design of unconventional magnets.

# INTRODUCTION

The exploration of magnetic materials with high storage density, ultrafast dynamics, and electric controllability is central to next-generation spintronic devices [1, 2]. In recent years, unconventional magnets that combine ferromagnetic-like responses with antiferromagnetic structures have emerged as promising candidates [3]. These systems host diverse emergent phenomena arising from two distinct physical sources: the non-relativistic magnetic geometry governed by exchange interactions and the relativistic spin-orbit coupling (SOC) effect. However, the textbook magnetic space group (MSG) framework assumes complete locking between lattice and spin spaces [4], which naturally conflates effects driven by magnetic geometry from those solely induced by SOC. Recently, the spin space group (SSG) theory has emerged as an extended framework to provide a comprehensice description of magnetic geometry by decoupling symmetry operations in lattice and spin spaces [5-16]. This has led to a new symmetry paradigm for understanding the physical effects of unconventional magnets, such as spin splitting, anomalous Hall effect, and magnetoelectric coupling [12, 17-28], i.e., employing SSG to describe the effects induced by magnetic order and MSG to characterize the SOC effects.

Despite this conceptual framework, its practical implementation for real materials remains hindered by the lack of a unified platform. In practice, different conventions between SSG and MSG frameworks, such as cell choices and coordinate systems, often hinder direct comparisons of physical responses across the non-relativistic and relativistic limits. Without a unified convention, reconciling these two symmetry descriptions requires manual realignment, which is inefficient and error-prone. While existing tools can identify SSG operations [29-31], they lack an automated workflow to bridge the SSG and MSG descriptions. Consequently, tracing the symmetry-breaking pathways of physical properties under SOC has remained largely manual, limiting systematic and high-throughput discovery of unconventional magnetism.

In this work, we address these challenges by introducing FINDSPINGROUP, an open-source platform for the automated identification of magnetic symmetry, magnetic phases, and various symmetry-constrained physical properties. Specifically, the platform is built on the recently developed oriented spin space group (OSSG) framework [32], which unifies the SSG and MSG frameworks and explicitly elucidates the symmetry-breaking pathways from SSG to MSG. FINDSPINGROUP generates standardized outputs, including magnetic cells, spin Wyckoff positions, spin site-symmetry groups, the spin Brillouin zone, and spin little co-groups. This automated workflow enables a systematic investigation of how various physical properties evolve as magnetic order and SOC are sequentially introduced, covering fundamental properties such as magnetizations, electric polarization, and spin polarization, as well as transport tensors, including the anomalous and nonlinear Hall effects. Beyond these capabilities, the practical utility of this workflow is demonstrated by its integration with other tools such as STensor [33] and Tensorsymmetry [34]. This interoperability, along with its widespread adoption [20, 21, 23, 32-37], establishes FINDSPINGROUP as a computational infrastructure for investigating unconventional magnets.

## SYMMETRY FRAMEWORK AND IMPLEMENTATION

### Spin space group and oriented spin space group

The symmetry operations of the SSG can be expressed as $\{g_s||g_l|\tau\}$, where $g_s = \{U_m(\varphi), TU_m(\varphi)\}$ consists of spin rotations $U_m(\varphi)$ and the time reversal $T$ that simultaneously reverses spin and momentum, while $g_l = \{C_n(\theta), PC_n(\theta)\}$ comprises lattice rotations $C_n(\theta)$ and the spatial inversion $P$. A magnetic structure is characterized by specifying the atomic positions (***r***) and magnetic moments (***m***). The action of an SSG operation on the tuple $(r, m)$ is defined as:

$$\{g_s||g_l|\tau\}(r, m) = (R(g_l)r + \tau, R(g_s)m) \qquad (1)$$

and the product of two SSG operations is defined as:

$$\{g_{s_1}||g_{l_1}|\tau_1\}\{g_{s_2}||g_{l_2}|\tau_2\} = \{g_{s_1}g_{s_2}||g_{l_1}g_{l_2}|g_{l_1}\tau_2 + \tau_1\} \qquad (2)$$

The general form of an SSG, denoted by $G_{SS}$, can be expressed as the direct product of a nontrivial SSG $G_{NS}$ and the spin-only group $G_{SO}$ [8]:

$$G_{SS} = G_{NS} \times G_{SO} \tag{3}$$

where $G_{NS}$ denotes the nontrivial SSG that every spin operation is combined with a spatial operation, except for the identity element, $G_{SO}$ consists of pure spin operations $\{g_s||E|0\}$ that describe the dimensionality of magnetic geometry. Based on the nomenclature of SSGs developed in Ref. [14], we extend the four-index nomenclature ($L_0.G_0.i_k.index$) of the nontrivial SSG $G_{NS}$ to Chen-Liu nomenclature ($G_0.L_0.i_k.index.L/P$) of the full SSG $G_{SS}$, where $G_0$, $L_0$, $i_k$, and $index$ represent the parent space group, sublattice space group, cell-expansion factor, and mapping sequence in the database, respectively. The additional fifth index $L$ or $P$ represents a collinear or coplanar spin-only group, respectively, and is omitted for non-coplanar configurations. The complete database of all SSGs, with their standard international notations, is available on the FINDSPINGROUP website.

When considering SOC effects, the operations in lattice and spin spaces become completely locked. The operations in the SSG that satisfy $U_m(\varphi) = C_n(\theta)$ constitute the corresponding MSG. Consequently, spin rotations are absorbed by the lattice rotations and are thus not explicitly expressed in MSGs. Although the SSG and MSG provide comprehensive descriptions of magnetic geometry and SOC effects, respectively, the arbitrary relative orientation between the spin and lattice spaces in the SSG [Fig. 1(a)] prevents a direct mapping to the MSG. This ambiguity hinders the comparative analysis of physical properties with and without SOC.

Historically, the characterization of phase transitions has relied on group-subgroup relationships, in which the symmetry-breaking path from a parent to a daughter phase is typically labeled by the irreducible representations of the parent group. For example, magnetic representations [38, 39] are adopted to characterize the symmetry descent from a paramagnetic space group to an MSG. However, the SSG possesses an intrinsic O(3) auto-equivalence in spin space, and its magnetic subgroups are obtained by selecting specific orientations. This unique property allows for a simultaneous description of the SSG and its MSG subgroup through the oriented SSG [32] by fixing the relative orientation between lattice and spin spaces [Fig. 1(b)], rather than introducing additional SSG representations.

Besides the alignment of spin-space operations of the nontrivial SSG with respect to the lattice, the OSSG introduces additional specification for spin-only groups in collinear and coplanar systems. The crystallographic direction [αβγ] is used to label the principal axis of the magnetic moment in collinear systems, $^{\infty}_{\alpha\beta\gamma}m1$, and the normal direction of the spin plane in coplanar systems $^{m}_{\alpha\beta\gamma}1$, respectively. Based on these, the connection between OSSG and the corresponding MSG can be established directly by locking the spin and lattice operations via the constraint $U_m(\varphi) = C_n(\theta)$ [Fig. 1(c)].

Take the OSSG of two altermagnets, CrSb and MnTe, as examples; they both share the OSSG $P^{-1}6_3/^{-1}m^1m^{-1}c\,^{\infty}_{\alpha\beta\gamma}m1$ (194.164.1.1.L), where the easy axis [αβγ] corresponds to [001] and [210], respectively. The corresponding MSG can be identified from completely locked operations. For CrSb, the OSSG corresponds to the MSG $P6_3'/m'm'c$, which can be generated from $\{-1 \cdot 6^1_{001} \| 6^1_{001}|\tau\}$, $\{-1 \cdot 2_{001} \| m_{001}|\tau\}$ and $\{-1 \cdot m_{210} \| m_{210}|\tau\}$, where $\tau = (0,0,1/2)$, For MnTe, the MSG $Cm'c'm$ can be obtained from $\{-1 \cdot m_{001} \| m_{001}|\tau\}$, $\{-1 \cdot 2_{210} \| m_{210}|\tau\}$,

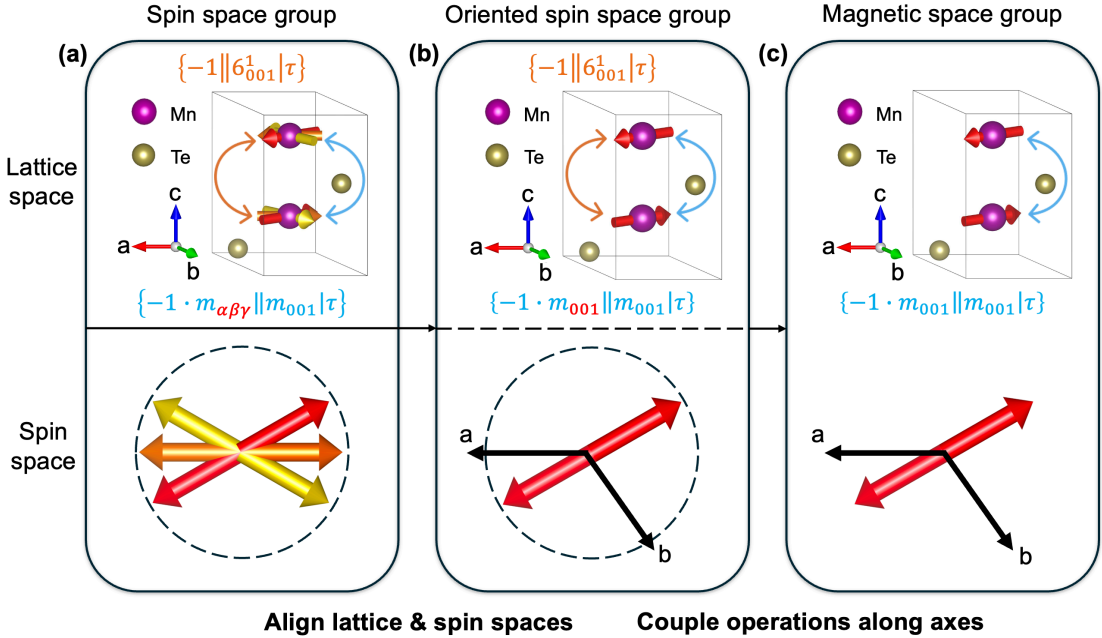

**Figure 1. Evolution from SSG to MSG via the oriented framework.** (a) In the SSG, the solid line in the middle separates two spaces, indicating that lattice and spin spaces are independent and have an arbitrary relative orientation. The multiple colored arrows and the dashed circle represent the orientational degeneracy in spin space. (b) In the OSSG, the dashed line indicates that the two spaces are aligned to define a material-

specific orientation of magnetic orders, but the operations remain decoupled. (c) Upon introducing SOC, the disappearance of the central line indicates complete coupling between the two spaces, and the magnetic moments are constrained along specific crystallographic axes. Only the subset of OSSG operations compatible with spin-lattice locking is preserved, yielding the corresponding MSG. The orientational degeneracy is removed, and the dashed circle therefore disappears. This construction integrates non-relativistic and relativistic descriptions, along with their symmetry evolution pathways, into a unified framework.

### *Identification and standardization*

Systematic identification of the OSSG requires a standard numerical framework to ensure consistency across the non-relativistic and relativistic regimes. In FINDSPINGROUP, the OSSG of a magnetic structure is determined by evaluating the symmetry operations within the direct product of the parent space group and the point group in spin space [14]. Throughout this work, the crystallographic part of the spin-group setting, including the basis vectors, cell setting, and origin choice, is expressed in the standard space-group convention of the Bilbao Crystallographic Server [40]. Besides, we introduce the .scif (spin crystallographic information file) format. Analogous to the .mcif format, the .scif format encodes the OSSG information as an extension to the underlying crystallographic structure. This approach preserves symmetry-resolved information across different settings and supports both tensor-symmetry analysis with STensor [33] and direct visualization of SSG symmetry in Jmol [41]. Detailed information on identification and visualization is provided in the Methods section.

Following the space groups and MSGs, the spin Wyckoff positions and spin site-symmetry groups are introduced to evaluate the Wyckoff-position splitting induced by magnetic order. Notably, the spin site-symmetry groups of an OSSG are always isomorphic to a subset of the 941 spin point groups. This isomorphism ensures that site-dependent effects can be exhaustively classified; for instance, only 375 spin point groups allow local magnetic moments. In reciprocal space, the OSSG dictates the symmetry of the spin Brillouin zone. While collinear and coplanar systems are characterized by the 24 centrosymmetric arithmetic crystal classes, noncoplanar SSGs correspond to all 73 arithmetic classes [42]. The construction of spin Wyckoff positions, the associated spin site-symmetry groups, and the little co-groups of wave vectors within the OSSG convention are discussed in Method. This ensures the precise identification of high-symmetry wavevectors, providing a comprehensive description of momentum-dependent spin physics.

## SYMMETRY-DICTATED PHYSICAL PROPERTIES

According to Neumann's principle, the symmetry of any physical property of a crystal must include the symmetry operations of its point group [43]. Consequently, crystallographic and magnetic point groups are often employed to impose symmetry constraints on physical properties such as ferroelectricity, magnetic moments, and various transport effects [44, 45]. However, unlike 32 point groups and 122 magnetic point groups, the 598 nontrivial spin point groups [46] and 941 spin point groups [10] are insufficient to provide an exhaustive description of SSGs. This is because SSGs permit arbitrary spin propagation $\{g_s||E|\tau\}$, where $g_s$ could be a non-crystallographic spin rotation. To bridge this gap, we directly employ the OSSG to impose symmetry constraints and trace the symmetry-breaking pathways as the magnetic geometry descends into the relativistic limit, where the locking of spin and lattice spaces can permit previously forbidden physical responses.

For a tensor *T*, the transformation under a symmetry operation of OSSG can be expressed as:

$$T'_{ijk\ldots} = \omega a_{il} a_{jm} a_{kn} \ldots T_{lmn\ldots} \tag{4}$$

where the indices $i, j, k, l, m, n, \ldots$ refer to Cartesian components *x*, *y*, and *z*. Following standard conventions [14], the *x*- and *z*-axes are aligned with the crystallographic *a*- and *c*-vectors to form a right-handed Cartesian system. The coefficients $a_{il}, a_{jm}, a_{kn} \ldots$ correspond to elements of the representation matrix derived from the symmetry operation $\{g_s||g_l|\tau\}$. The factor $\omega$ accounts for the parity under spatial inversion and time reversal.

Based on whether they change sign under spatial inversion and time reversal, tensors are classified as either polar or axial, and as time-dependent and time-independent, respectively. Therefore, $\omega$ is determined by the determinants of the symmetry operations in lattice and spin spaces:

$$\omega = \begin{cases} 1 & \text{time-dependent polar vector} \\ det(D(g_l)) & \text{time-dependent axial vector} \\ det(D(g_s)) & \text{time-independent polar vector} \\ det(D(g_l))det(D(g_s)) & \text{time-independent axial vector} \end{cases} \quad (5)$$

where $det(D(g_l))$ and $det(D(g_l))$ denote the determinant of the operations in lattice and spin spaces, respectively. Improper rotations yield a determinant of -1 and correspond to rotations combined with spatial inversion in lattice space or time reversal in spin space.

The overall workflow of FINDSPINGROUP is illustrated in Fig. 2. Starting from structure files (.cif, .mcif, .scif), the program identifies the OSSG, assigns the corresponding standardized database index, and derives outputs such as conventional and primitive magnetic cells, spin Wyckoff positions, and the spin Brillouin zone. By applying the tensor constraints defined in Eq. (4), FINDSPINGROUP evaluates a range of physical properties, including magnetization, electric polarization, spin splitting, and various transport effects. Representative applications to these classes of physical properties are discussed below.

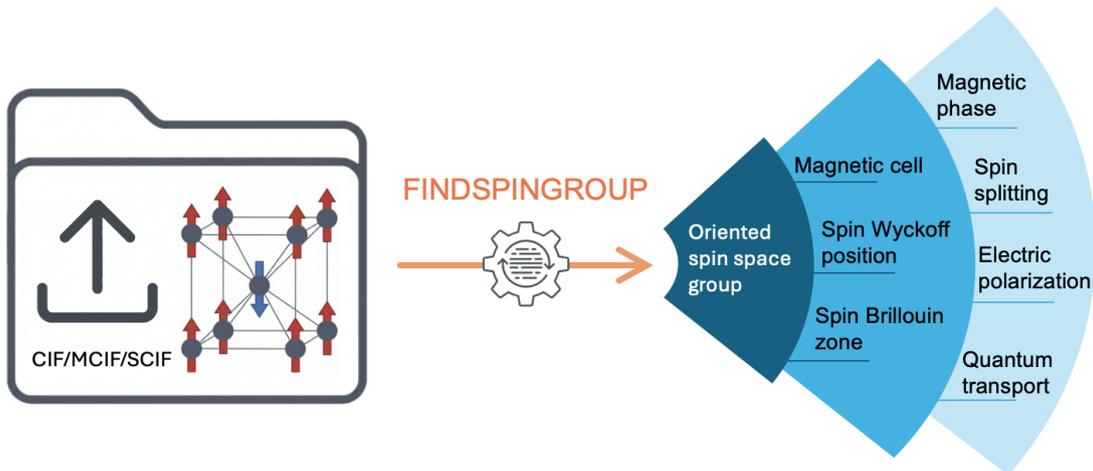

**Figure 2. Overview of the workflow of FINDSPINGROUP.** Starting from magnetic-

structure input in CIF, MCIF, or SCIF format, FINDSPINGROUP identifies the OSSG and uses it as the working convention for downstream analysis. Within this unified framework, the platform determines the magnetic cell, spin Wyckoff positions, and spin Brillouin zone, and further evaluates symmetry constraints on magnetic phase, spin splitting, electric polarization, and quantum transport in the presence or absence of spin-orbit coupling.

*Magnetization and electric polarization*

As discussed before, the OSSG framework, which reconciles the SSG description of the phenomenological dichotomy between ferromagnetism and antiferromagnetism [32] with the MSG description of SOC effects, is employed to design time-reversal-odd physical responses in unconventional magnets. Below, we begin with the net magnetization $M$, a time-dependent axial vector, to show how the OSSG establishes a symmetry classification of magnetism.

Symmetry analysis can be applied separately to the spin magnetization $M^S$ and the orbital magnetization $M^O$ in the nonrelativistic limit. Although both are time-dependent axial vectors, $M^S$ is governed by the spin-space operation $g_s$ while $M^O$ is constrained by the lattice-space operation $g_l$ and the time reversal component in $g_s$. Specifically, the spin-space point group, formed by all spin-space operations in OSSG, constrains $M^S$. In contrast, $M^O$ is constrained by the effective magnetic point group, which is obtained by mapping the operation $\{g_s||g_l|\tau\}$ to $g_l$ and $Tg_l$ for proper and improper spin-space operation $g_s$, respectively. The respective transformation relations are expressed as:

$$M_i^S = det(D(g_s)) \sum_m D(g_s)_{il} M_l^S, \tag{6}$$

$$M_i^O = det(D(g_s)) \sum_m D(g_l)_{il} M_l^O. \tag{7}$$

This distinction leads to different physical consequences. The symmetry-protected zero $M^S$ defines the boundary between ferromagnets/ferrimagnets (FMs/FiMs) and antiferromagnets (AFMs) [32], serving as a prerequisite for exploring unconventional magnets with time-reversal-odd physical responses. On the other side, the spin-mirror operation in collinear and coplanar magnets enforces a zero $M^O$ in the absence of SOC, while non-coplanar magnets can support a nonzero $M^O$ originating from the non-relativistic magnetic geometry. Furthermore, the symmetry requirements for $M^O$ are identical to those of the anomalous Hall conductivity, which corresponds to the antisymmetric part of the conductivity tensor and thus an axial pseudovector [47]. This implies that a magnetic-geometry-induced AHE can only emerge in non-coplanar magnets [48, 49]. In contrast, for collinear and coplanar systems, the AHE originates solely from relativistic SOC effects.

The classification procedure of magnetic phases under OSSGs is summarized in Fig. 3. A polar spin-space point group, defined as a subgroup of $\infty m$, indicates a nonzero $M^S$. Systems with a nonzero $M^S$ are further distinguished by their spin Wyckoff positions (SWPs), of which a single type identifies an FM and multiple inequivalent types indicate a FiM. In the case of FiM, when the magnetization vanishes, the system is further classified as a compensated ferrimagnet (cFiM). Conversely, nonpolar spin-space point groups correspond to AFMs. Unlike FMs, which maintain a net magnetization in both non-relativistic and relativistic limits, AFMs are further distinguished by their response to SOC. If there exists a net magnetization induced by SOC effects in an AFM, i.e., allowed by the fully coupled operations in its OSSG, it is further classified as spin-orbit magnetism (SOM) [32]. Such a unique class of AFM manifests FM-like properties related to finite magnetization, such as the anomalous Hall effect, anomalous Nernst effect, and magneto-optical Kerr effect, etc.

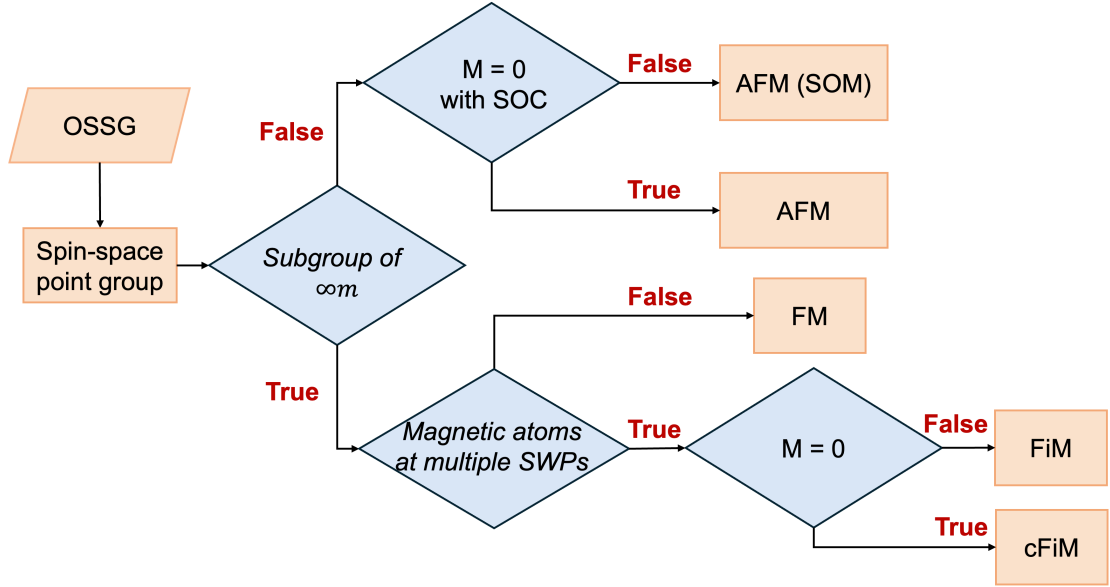

**Figure 3. Symmetry classification of magnetic phases.** The diagram summarizes the classification route from the OSSG, the associated spin-space point group, the SWPs, and the net magnetization M, yielding AFM, SOM from an antiferromagnetic state [AFM(SOM)], FM, FiM, or cFiM.

Beyond magnetic classification, the OSSG framework also provides a fundamental basis for studying coexisting ferroic orders and their correlations, which are essential for understanding multiferroic couplings in unconventional magnets. For example, this analysis extends naturally to the macroscopic electric polarization [23-25], which is a time-independent polar vector. In addition to conventional lattice-driven ferroelectricity, the OSSG distinguishes two distinct physical origins of ferroelectricity related to magnetic order. In a non-polar lattice, spin-induced ferroelectricity occurs when magnetic ordering breaks spatial symmetry into a polar group. In comparison, SOC-induced ferroelectricity is identified when a polar group forms upon the inclusion of SOC. By evaluating these symmetry constraints, FINDSPINGROUP determines whether the ferroelectric state is driven by lattice order, magnetic order, or SOC effects. Distinguishing these mechanisms facilitates the exploration of magnetoelectric coupling in unconventional magnets, thereby enabling the design of non-volatile, low-power, and high-efficiency spintronic devices.

*Spin splitting and spin polarization*

Besides macroscopic properties, the OSSG evaluates how magnetic geometry governs the spin-polarized band structure in reciprocal space, where the wave vector *k* is treated as a time-dependent axial vector and remains invariant under proper spin rotations. In FMs, FiMs, and cFiMs, the spin-space point group is polar, leading to Zeeman-type splitting in such systems. In AFMs, the SSG protects a vanishing net magnetization, corresponding to the absence of spin polarization at the Γ point in momentum space. Away from Γ, when the spin-space part of the *k*-little co-group is polar, the momentum-dependent spin splitting is allowed. This condition requires breaking the combination of spatial inversion and time reversal (*PT*), which is also an MSG operation. In addition, such spin splitting requires the absence of the $D_\infty$ or $D_2$ group in spin space, which could occur in collinear and noncollinear magnetic configurations, respectively [14]. This analysis also distinguishes between spin-polarization components at any *k*-point induced by magnetic geometry and the additional relativistic contributions that emerge only upon the introduction of SOC. For collinear AFMs that support momentum-dependent spin splitting without SOC, FINDSPINGROUP explicitly assigns the label "altermagnet" [12].

This method can also be applied to two-dimensional spin-split AFMs. Compared with three-dimensional cases, a prerequisite step is to identify the orientation of the non-periodic direction of the structural input and then to apply constraints on the spin polarization. In practice, the three-dimensional OSSG is first determined. By removing the *k*-point components along the non-periodic direction, the two-dimensional spin Brillouin zone, high-symmetry *k*-points, and spin little co-groups for any *k*-points are identified. The reduction in dimensionality can induce additional symmetry-protected degeneracies, i.e., the spin polarization at general *k* points can be further constrained by improper spin rotation coupled with lattice rotation $C_{2z}$, or proper spin rotations associated with the mirror operation $M_z$ [50, 51].

## **EXAMPLES**

To illustrate the functionality of our algorithm and program, we present several examples in which FINDSPINGROUP is used to identify OSSGs and analyze their associated physical properties, both in the SOC-free and SOC-included cases.

### *Altermagnet $V_2Se_2O$*

The full symmetry of two-dimensional (2D) altermagnet $V_2Se_2O$ (Fig. 4(a)) [19, 52, 53] is described by the OSSG $P^{-1}4/^1m^1m^{-1}m^{\infty\alpha\beta\gamma}m1$ (123.47.1.1.L), where [αβγ] denotes the easy axis under SOC. This system is classified as an AFM due to its spin-space point group ∞/mm (Fig. 4(c)). In the absence of SOC, this OSSG belongs to the arithmetic crystal class 4/mmm$P$, the corresponding reduced spin Brillouin zone is constructed with all high-symmetric *k*-points identified by the red triangle in Fig. 4(b). Symmetry analysis reveals that this collinear AFM is an altermagnet in the SOC-free limit, characterized by momentum-dependent spin splitting. The spin polarization consists of only the $S_x$ component, while the electronic bands along the Γ-M direction remain spin degenerate due to the $\{2_{010}\|m_{1-10}\}$ symmetry operation. Furthermore, the $\{2_{010}\|4_{001}\}$ operation enforces a sign reversal of the spin texture under a four-fold lattice rotation in momentum space, thereby dictating the *d*-wave spin texture and the spin-polarized Fermi surfaces as shown in Fig. 4(d).

Upon the introduction of SOC, the [100] easy axis breaks the four-fold lattice rotation symmetry, yielding an OSSG subgroup that is isomorphic to the MSG *Pmm′m′*. As a result, the system is further classified as an SOM. Notably, a nonzero net magnetization emerges parallel to the Néel vector, which is distinct from the conventional ferromagnetic canting induced by Dzyaloshinskii-Moriya interactions, where the net magnetization is perpendicular to the Néel vector. This effect arises from SOC-induced Wyckoff splitting, in which the symmetry equivalence between two opposite-spin sublattices is lifted. The two V atoms at spin Wyckoff position 2*f* are split into two different magnetic Wyckoff positions, 1*c* and 1*e*.

The symmetry breaking also leads to an expansion of the reduced Brillouin zone where the high-symmetry Γ-M path reduces to a set of general *k*-points (Fig. 4(b)). This reduction lifts spin degeneracy throughout the whole Brillouin zone, allowing the emergence of spin polarization $S_x$. Although the $\{2_{100}\|m_{100}\}$ symmetry ensures that the spin polarization at all high-symmetry points remains restricted to the $S_x$ component, the further reduction of symmetry enables the emergence of an additional $S_y$ component at general k-points, while $S_z$ remains strictly forbidden by the $\{m_{001}\|2_{001}\}$ symmetry. While MSG permits a uniform $S_x$ distribution across the Brillouin zone, the d-wave character remains dominant because of the negligible SOC (Fig. 4(d)). The small $S_y$ projection observed in the DFT calculations is also consistent with the symmetry analysis. This evolution explains how the OSSG captures the symmetry-breaking pathway and the resulting modulation of momentum-dependent spin textures, providing a microscopic foundation for the design of tunneling magnetoresistance, piezomagnetism, and spin-splitting torque.

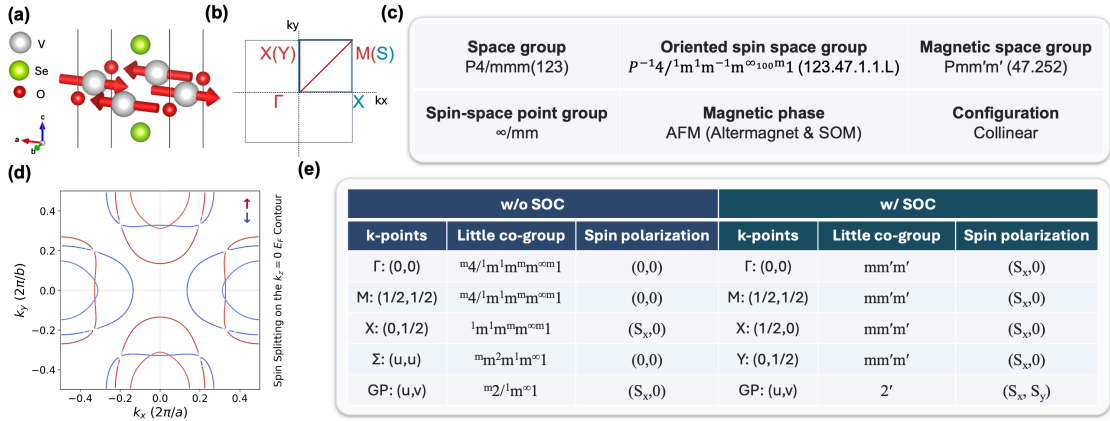

**Figure 4. Symmetry analysis of 2D altermagnet V$_2$Se$_2$O.** (a) Magnetic structures, (b) spin Brillouin zone, and (c) group information of 2D altermagnet V$_2$Se$_2$O. The area around the red lines is the reduced BZ of SSG, and the blue one is the reduced BZ of MSG. (d) DFT-calculated spin splitting on the $k_z = 0$ plane at the Fermi level in the absence of SOC. (e) Momentum-dependent spin splitting in V$_2$Se$_2$O without and with SOC. All the information except panel (d) can be output by FINDSPINGROUP.

## *Spin-induced ferroelectricity in altermagnet MnSe$_2$*

The OSSG framework implemented in FINDSPINGROUP also facilitates the identification of coexisting ferroic orders and the exploration of magnetic switch mechanisms. As stated before, this approach effectively distinguishes between spin-induced and SOC-induced ferroelectricity, enabling the electric control of unconventional magnets [23-26]. For example, the non-magnetic state of $MnSe_2$ [54] crystallizes in the centrosymmetric space group *Pa*-3, which forbids spontaneous electric polarization. However, the emergence of a supercell antiferromagnetic order leads to a symmetry reduction as illustrated in Figs. 5(a) and 5(b). Symmetry analysis indicates that the spatial part of the OSSG is *Pca*$2_1$, a polar group that permits a spontaneous electric polarization along *z* direction. This mechanism differs from traditional multiferroics governed by exchange striction models, such as $MnS_2$ and $Cu_2MnSnS_4$, where spatial inversion is broken while the T$\tau$ symmetry is preserved, resulting in hidden altermagnets [55, 56]. In contrast, the magnetic ordering in $MnSe_2$ simultaneously breaks both, inducing a macroscopic ferroelectric state and enabling momentum-dependent spin splitting.

Beyond the coexistence of ferroic orders, FINDSPINGROUP enables the systematic identification of switching trajectories between symmetry-related domain states. By determining the coset representatives of the space group relative to the OSSG, 36 potential switching paths are identified. Subsequent calculations can be employed to identify the trajectories with the lowest energy barriers. For instance, considering the operations $\{1\|-1|0,0.5,0\}$ and $\{-1\|-1|0,0.5,0\}$ (Figs. 5(d) and 5(e)), the former represents a path where the spin polarization remains invariant during ferroelectric switching, whereas the latter corresponds to the simultaneous switch of electric polarization and spin splitting. Such symmetry analysis facilitates the design of spin-induced ferroelectricity and the prediction of whether the ferroelectric state can control coexisting FM-like responses. By distinguishing between this exchange-driven mechanism and the more common SOC-induced ferroelectricity, the OSSG framework provides a microscopic roadmap for designing multiferroic materials where electric and magnetic degrees of freedom are coupled.

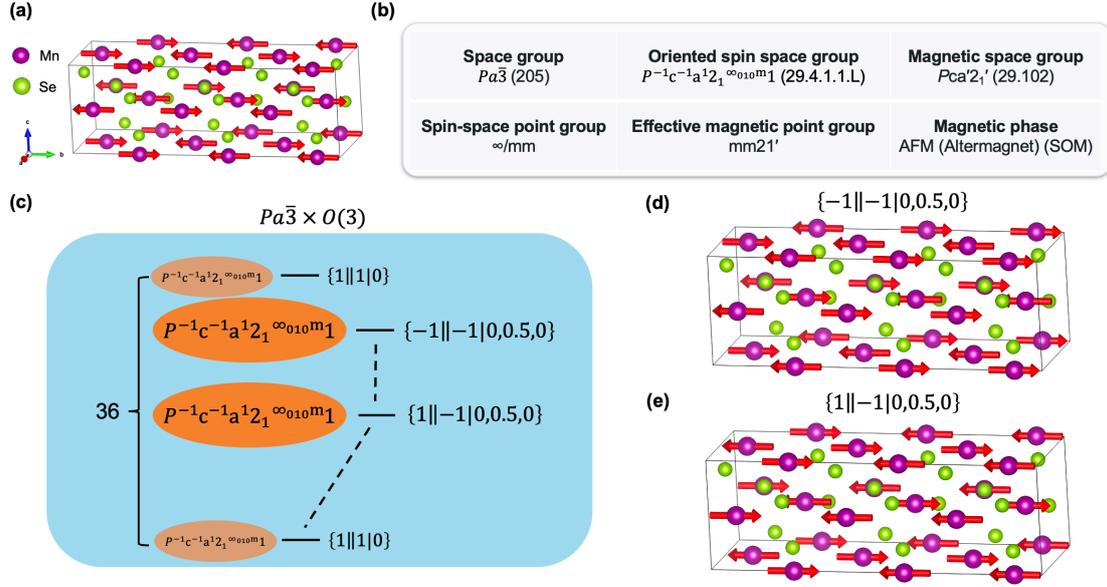

Figure 5. **Symmetry analysis of ferroelectric altermagnet MnSe$_2$.** (a) Magnetic structures and (b) group information of altermagnet MnSe$_2$. (c) Schematic illustration of the coset decomposition of the spin space group as a subgroup of the direct product of the space group $Pa\bar{3}$ and $O(3)$, corresponding to 36 possible magnetic phase transition pathways. Panels (d) and (e) represent the magnetic phases obtained via operations $\{-1\|-1|0,0.5,0\}$ and $\{1\|-1|0,0.5,0\}$ from (a), respectively.

## *All-in-all-out antiferromagnet CoNb$_3$S$_6$*

The all-in-all-out AFM CoNb$_3$S$_6$ (Fig. 6(a)) represents a typical case where OSSG captures the coexistence of spin degenerate bands and the magnetic geometry induced AHE. It is described by the OSSG $P^{3^2_{001}}6_3{}^{m_{100}}2^{m_{010}}2|(2_{12\text{-}\gamma}, 2_{21\gamma}, 1)$ (182.4.4.2). A distinct feature of this OSSG is the coexistence of a hexagonal space group P622 with a tetrahedral spin-space point group -43m, whose point-group components are incompatible (Fig. 6(b)). Such a configuration goes beyond the description of 941 spin point groups and can only be characterized by the SSG or OSSG. This example demonstrates that spin point groups alone cannot exhaustively characterize the physical properties of magnetic geometry, unlike classification using conventional point groups and magnetic point groups.

Despite the absence of *PT* symmetry, the spin little co-group at the general *k*-point is $^{222}1$, which strictly forbids the spin polarization throughout the whole Brillouin zone [14, 16, 57, 58]. However, while the electronic band structure of CoNb$_3$S$_6$ remains spin degenerate in the absence of SOC, it supports a nonzero SOC-free AHE since its OSSG corresponds to the ferromagnetic effective magnetic point group $62'2'$ (Fig. 6(c) and 6(d)). This AHE originates from the non-coplanar magnetic geometry, while such a response is forbidden in collinear and coplanar AFMs due to the presence of effective time reversal symmetry from the spin-only group. The coexistence of spin degeneracy and the AHE indicates that the physical origins and symmetry requirements of these two effects are distinct [3].

When SOC is included, the OSSG corresponds to the chiral MSG $P32'$, which allows the emergence of a net magnetization along the [001] direction and reclassifies the magnetic phase as SOM. Furthermore, SOC breaks the spin-translation group $^{222}1$, leading to the emergence of SOC-induced spin polarization as shown in Fig. 6(e). Specifically, the presence of the $3^1_{001}$ operation dictates that, for the Γ-A and H-K paths, only the $S_z$ component is permitted. In contrast, on the $k_y = 0$ plane, the presence of the $T2_{100}$ symmetry allows for the $S_y$ and $S_z$ components, while at general $k$-points, spin polarization in arbitrary directions is permitted.

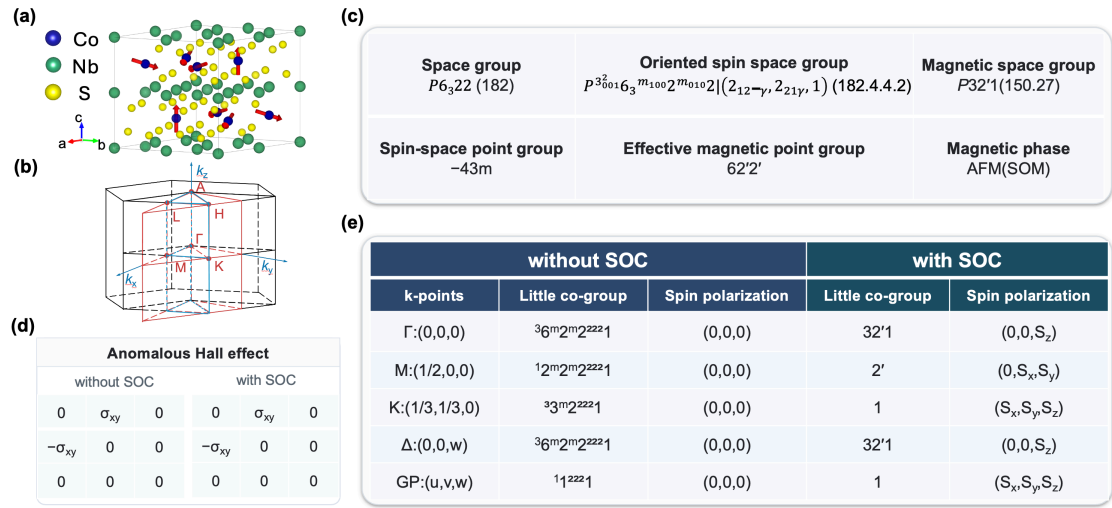

**Figure 6. Symmetry analysis of all-in-all-out antiferromagnet CoNb$_3$S$_6$.** (a) Magnetic structures, (b) spin Brillouin zones (blue: SSG-reduced BZ; red: MSG-reduced BZ), and (c) group information, (d) anomalous Hall conductivity without and with SOC, and (e) momentum-dependent spin splitting of CoNb$_3$S$_6$. All the information

can be output by FINDSPINGROUP.

## SUMMARY

The FINDSPINGROUP platform provides a systematic computational framework for symmetry analysis and the exploration of physical properties in unconventional magnets. By implementing the OSSG framework, it treats magnetic geometry and SOC effects within a single workflow and combines standardized structural conventions, spin site-symmetry groups, spin Brillouin zones, and spin little co-groups. Based on the OSSG framework, the platform enables automated tracking of symmetry-breaking pathways from the non-relativistic to the relativistic limit, facilitating the symmetry classification of magnetic phases and the disentanglement of physical origins for emergent phenomena such as spin splitting and the AHE. This scheme provides a clear physical picture for understanding the interplay between magnetic geometry and SOC. Furthermore, the introduction of the SCIF format standardizes spin-group symmetry analysis in a manner analogous to conventional crystal structure analysis, enhancing the efficiency of high-throughput screening and design of unconventional magnets. Overall, the FINDSPINGROUP platform offers physical navigation for the high-throughput discovery of unconventional magnets, facilitating the future development of spintronic devices.

## METHODS

### *Identification, equivalence and index*

From the input magnetic structure, we determine a magnetic primitive cell and the crystallographic operations preserving it using spglib [59]. The site moments are then used to classify the magnetic configuration as collinear, coplanar, or non-coplanar and to determine the admissible spin-space operations. The identified symmetry is the set of all operations $\{g_s||g_l|\tau\}$ that leave the magnetic structure invariant, with action (1).

Classification is performed on the equivalence class of this operation set under orientation-preserving affine changes of real-space coordinates, together with proper rotations of spin space. Writing an affine transformation as $(A, a) \in \text{Aff}^+(3)$ and a spin-space rotation as $Q \in \text{SO}(3)$, two SSG descriptions are equivalent if their operations are related by conjugation,

$$\{g_s \| g_l | \tau\} \rightarrow \{Q g_s Q^{-1} \| A g_l A^{-1} | A\tau + a - A g_l A^{-1} a\} \qquad (8)$$

This removes basis, origin, and spin-frame choices while preserving the magnetic symmetry.

The nomenclature is assigned only after reduction to the canonical database representative of this equivalence class at database.findspingroup.com. Thus, the reported label identifies an equivalence class rather than a raw input-dependent realization. For certain coplanar families, especially those associated with nontrivial spin-space point group 2 or 222, the coarse suffix P does not uniquely specify the full group $G_{SS} = G_{NS} \times G_{SO}$. Although the nontrivial part $G_{NS}$ is fixed, extending it by the coplanar spin-only part $G_{SO}$ can yield, under the equivalence defined above, up to three inequivalent full-group representatives. These refined coplanar representatives are distinguished in the database by the suffixes P1, P2, and P3.

### *Spin Wyckoff positions and spin Brillouin zones*

Spin Wyckoff positions are obtained as site orbits under the OSSG. For a representative site, the associated spin site-symmetry group is the subgroup of OSSG operations that leaves the site invariant, and it determines the symmetry-allowed moment components at that site. This fixes the splitting chain from the parent space group to the OSSG and then to the corresponding MSG in one convention.

The spin Brillouin zone is constructed from the effective reciprocal-space action of the OSSG. For an OSSG operation $\{g_s \| g_l | \tau\}$, the induced action on a wave vector is

$$\{g_s \| g_l | \tau\} : k \rightarrow \det(D(g_s)) \, ((D(g_l))^{-1})^T k \qquad (9)$$

modulo reciprocal lattice vectors. The resulting effective k-space operations are reduced to an arithmetic crystal class, which fixes the spin Brillouin-zone type, the standardized primitive transformation and the tabulated high-symmetry k points. For a given wave vector k, the spin little group is obtained from the subset of OSSG operations that preserves k under the same action, modulo reciprocal lattice translations. The corresponding spin little co-group is defined by removing the translational part of this little group. In this way, the spin Brillouin zone, its labeled high-symmetry points, and the associated spin little co-groups and magnetic little co-groups are determined in the same oriented convention as the real-space symmetry.

## *SCIF representation*

To record information on magnetic materials and their OSSG symmetries, we introduce the standardized .scif (spin crystallographic information file) format for depositing and disseminating SSG-related data. Similar to .cif and .mcif, the .scif file records the spin space group assignment together with the underlying crystallographic structure of a material, while allowing symmetry information to be stored and transformed across different conventions and settings. It further provides symmetry-resolved information required for subsequent spin-group-based property calculations, which can be directly accessed by tools such as STensor [33] for tensor-symmetry analysis. In addition, .scif enables direct visualization of SSG symmetry with compatible tools such as Jmol [41], facilitating intuitive inspection and verification.

## *Tolerances parameters*

The default tolerance parameters of the current implementation are used throughout: space_tol = 0.02 Å, mtol = 0.02 $\mu_B$, meigtol = 2 × $10^{-5}$, matrix_tol = 0.01, and parser_atol = 0.02. Here, space_tol is the real-space distance tolerance used for crystallographic symmetry matching, space-group determination, and standard-setting construction; mtol is the magnetic-moment tolerance used for moment matching and for classifying the magnetic configuration as collinear, coplanar, or non-coplanar; meigtol is the eigenvalue tolerance used in point-group candidate construction; matrix_tol is the tolerance used to compare operation matrices during standardization and equivalence reduction; and parser_atol is the absolute tolerance used in SCIF expansion to check whether symmetry-generated moments assigned to the same crystallographic site remain numerically consistent.

*DFT calculations*

All DFT calculations were performed using the projector augmented-wave method as implemented in the Vienna *ab initio* simulation package [60, 61]. The generalized gradient approximation of Perdew-Burke-Ernzerhof-type exchange-correlation potential [62] was adopted. A plane-wave cutoff energy of 520 eV and an electronic convergence criterion of $10^{-7}$ eV were used. The self-consistent calculations were performed with a Γ-centered 15×15×1 k-mesh, and the spin texture was calculated with a Γ-centered 101×101×1 mesh.

# CODE AVAILABILITY

FINDSPINGROUP is an open-source platform, with the source code available at https://github.com/LiuQH-lab/FindSpinGroup.

# FUNDING

This work was supported by National Natural Science Foundation of China under Grant Nos. 12525410, 12274194, 12574275 and 12534003, Guangdong Provincial Quantum Science Strategic Initiative under Grant No. GDZX2401002, Guangdong Provincial Key Laboratory for Computational Science and Material Design under Grant No. 2019B030301001, Shenzhen Science and Technology Program (Grant No. RCJC20221008092722009 and No. 20231117091158001), the Innovative Team of General Higher Educational Institutes in Guangdong Province (Grant No. 2020KCXTD001), the Open Fund of the State Key Laboratory of Spintronics Devices and Technologies (Grant No. SPL-2407) and Center for Computational Science and Engineering of Southern University of Science and Technology.

# AUTHOR CONTRIBUTIONS

Q.L. conceived the project. Y.Y. and X.C. developed the theoretical framework and core codebase and built the database. Y.Z. developed the spin-space-group-labeling code.

Y.H.L. calculated and curated the spin-Wyckoff-position data. R.X. calculated and curated the spin-space-group operations data. J.L. provided theoretical guidance on higher-order nonlinear tensors. Y.T.L. provided guidance on magnetic classification. Y.Y., X.C., and Q.L. wrote the manuscript. All authors discussed the results and approved the manuscript.

***Conflict of interest statement.*** None declared.